\documentclass[smallextended,natbib]{svjour3}
\usepackage{graphicx,picins,wrapfig}           
\usepackage[ngerman,english]{babel}
\usepackage[T1]{fontenc}
\usepackage[ansinew]{inputenc}
\usepackage{floatflt}           
\usepackage{pifont,geometry,fleqn,hyperref}
\usepackage{placeins}
\usepackage{ae}
\usepackage{amsmath,amssymb,amstext,amscd}
\usepackage{bm}
\usepackage{txfonts}
\usepackage{natbib}
\usepackage[font=scriptsize]{subfig}            
\usepackage{times}
\usepackage{lettrine}
\usepackage{fancyhdr}           
\usepackage{deluxetable}
\newcommand\aj{AJ~}
\newcommand\apj{ApJ~}
\newcommand\mnras{MNRAS~}
\newcommand\aap{A\&A~}

\newcommand\apjl{ApJL~}

\newcommand\nat{Nature~}

\defcitealias{helm99a}{H99}
\newcommand{\ocen}{$\omega$\,Cen\:}

\begin{document}

\title{Halo Streams in the Solar Neighborhood}
\author{Rainer J. Klement}
\email{klement@mpia.de}
\institute{Max-Planck-Institut f\"ur Astronomie, K\"onigstuhl 17, D-69117 Heidelberg}
\begin{abstract}
The phase-space structure of our Galaxy holds the key to understand and reconstruct its formation. The $\Lambda$CDM model predicts a richly structured phase-space distribution of dark matter and (halo) stars, consisting of streams of particles torn from their progenitors during the process of hierarchical merging. While such streams quickly loose their spatial coherence in the process of phase mixing, the individual stars keep their common origin imprinted into their kinematic and chemical properties, allowing the recovery of the Galaxy's individual ``building blocks''. The field of Galactic Archeology has witnessed a dramatic boost over the last decade, thanks to the increasing quality and size of available data sets. This is especially true for the solar neighborhood, a volume of 1-2 kpc around the sun, where large scale surveys like SDSS/SEGUE continue to reveal the full 6D phase-space information of thousands of halo stars. In this review, I summarize the discoveries of stellar halo streams made so far and give a theoretical overview over the search strategies imployed. This paper is intended as an introduction to researchers new to field, but also as a reference illustrating the achievements made so far. I conclude that disentangling the individual fragments from which the Milky Way was built requires more precise data that will ultimately be delivered by the \textit{Gaia} mission. 
\end{abstract}

\section{Introduction}
Not until the past two decades have we begun to understand the composition of our Galaxy in a universal context of galaxy formation and evolution, thanks to great progress in both observations and theoretical work. Historically, there have been two different models for the early formation of the Milky Way's stellar halo. \citet{egg62} found strong correlations between the metallicity and eccentricity of stars in the Solar neighborhood. To explain the highly eccentric orbits of the oldest stars, they argued that at the time of formation of these stars, the Galaxy could not have been in dynamical equilibrium, but in a state of rapid gravitational collapse ($\sim10^8$ yr) from a larger homogeneous spheroid, the ``protogalaxy''. After the Galaxy reached dynamical equilibrium and became rotationally supported, further star formation would have taken place in a metal-enriched disk, thereby explaining the disk-like orbits of the metal-rich stars. This scenario is called a \textit{monolithic collapse}. Although the correlation between eccentricity and metallicity found by \citet{egg62} is empirically correct for samples including both disk and halo stars, many authors later questioned its validity for a pure halo sample and it was definitively discarded from a kinematically unbiased sample of 1203 metal-poor stars \citep[][and references therein]{chi00}.

In 1978, \citeauthor{searle78} published their seminal study on the distribution of globular clusters in the outer Milky Way halo; they found no abundance gradient, but a substantial spread in the ages of the clusters. They concluded that, although the lack of an abundance gradient would still be consistent with cluster formation during the rapid free-falling phase of a galactic collapse (because the freefall timescale is much shorter than the enrichment timescale), globular clusters would not have such a high internal chemical homogeneity, if they would have formed during a monolithic collapse. Therefore, \citet{searle78} proposed a slow ($\gtrsim10^9$ yr) and more chaotic process in which the Milky Way was gradually built through the merging of several small protogalactic ``fragments''. This is called a \textit{bottom-up scenario}, in which galaxies are generally formed through the amalgamation of smaller fragments. However, \citet{chi00} found that the density distribution of the halo changes from highly flattened in the inner parts to nearly spherical in the outer parts with no discrete boundary. Also they found a continuous negative rotational velocity gradient with height above the Galactic plane, both results that are hard to arrange with the original \citet{searle78} scenario, because chaotic merging would not produce such kinematic and spatial structures. 

Therefore both simplistic models cannot fully explain the distribution, kinematics and abundances of halo stars in the Milky Way. However, the stellar halo only accounts for $\sim1\%$ of all stars in the Milky Way, and a more general approach towards understanding the formation of the Milky Way, and disk galaxies in general, is needed. Such an approach has to take into account cosmological principles and the nature and properties of dark matter. Dark matter is the dominant form of matter in the Universe and it should play the key role in galaxy formation through gravitational collapse \citep{press74, white78}. Numerical and semi-analytic models show that, if the random velocities of the dark matter particles are small compared to the speed of light (cold dark matter or CDM), present day galaxies can be formed through gravitational collapse of the small density fluctuations which where present in the early Universe after the big bang. While the average density of the cosmic matter declined as the Universe expanded, some density enhancements of sufficient size became more pronounced, because of their own gravitational attraction. These protosystems drew in more matter from surrounding regions and subsequently merged to larger and larger structures, increasing further the lumpiness of the once highly (but not perfectly) uniform distribution of matter. The ``trunks'' of such ``merger trees'' in the \textit{hierarchical scenario} are the present day galaxies \citep[e.g.][]{katz92,cole94,mo98,stei99,bek01,stei02}. Although these simulations lack high enough resolution to study the detailed spatial and kinematic distribution of stars, they qualitatively can explain the formation and properties of different galactic components. Recent simulations of galaxy formation in the $\Lambda$CDM cosmology were able to build Milky Way-like spiral galaxies including thin disk, thick disk, bulge and halo \citep{abadi03a,abadi03b,brook04,sam04,gov07}. 

The hierarchical merging scenario predicts a richly structured phase-space distribution of dark matter and stars \citep[e.g.][]{bul05}. Early evidence for remnants of accretion events in the outer halo was found by, e.g., \citet[][five stars at $\langle d\rangle\sim5$ kpc distance from the sun]{doi89}, \citet[][four stars at $d\sim30$ kpc]{arn92}, \citet[][discovery of the disrupting Sagittarius dwarf galaxy as a clump in radial velocity space of several 100 K and M giants]{iba94} or \citet[][several metallicity/ phase-space clumps in a sample of 154 stars at $d\lesssim8$ kpc]{maj96}. The last years have seen a dramatic increase in the detection of halo substructure thanks to automatic large-scale photometric surveys such as 2MASS \citep{skr06} and SDSS \citep{sto02}. Prominent examples are the currently accreted and disrupted Sagittarius dwarf galaxy with its associated tidal tails \citep{iba01,maj03,bel06a}, the Orphan stream \citep{bel06a}, the Virgo overdensity \citep{viv01,new02,jur08} or the globular clusters Palomar 5 \citep{ode01,grill06} and NGC 5466 \citep{bel06b,gri06}.

Numerical simulations predict that most accreted satellites would also spread their tidal debris on eccentric orbits into the Milky Way's inner halo \citep{helm99b} or even the disk component \citep{abadi03b,meza05}; thus we can expect to find such ``fossil remnants'' right in our immediate surroundings. Although no longer spatially coherent, such stellar streams keep their common origin imprinted into their chemical and dynamical properties. This allows us to study the formation history of our Galaxy by disentangling its individual ``building blocks'' from the phase-space distribution of solar neighborhood halo stars, for which we have much better kinematic and abundance informations than for their distant counterparts. The difficulty, however, lies in the very low density contrast that is expected for individual streams. Due to the short dynamical time scales near the disk, a single progenitor can dispose multiple streams in the solar neighborhood in a short period of time, leading to an expected total of several hundreds of cold ($\sigma\lesssim5$ km s$^{-1}$) stellar streams \citep{helm99b,helm03,gou03}\footnote{Compare this number to an expected number of tens of thousands of CDM particle streams \citep{helm03,vog08}.}. For obtaining most of the informations hidden in the phase-space structure of local stars it follows that (i) a large sample of thousands of halo stars with very accurate velocities would be needed and (ii) the strategies for finding the individual streams and deriving informations about the past should be optimized. The first point is currently addressed by large-scale surveys such as RAVE \citep{stei06} and SEGUE \citep{yan09} and will be completely satisfied when the \textit{Gaia} mission \citep{per01} is completed. The second point is part of this review, which should be seen both as an introduction for people new to this field as well as a reference summarizing the current state of research. 

\section{The formation of stellar halo streams\label{sec:s2}}
In this section I will summarize the main processes that lead to the formation of stellar halo streams that might show up in samples of nearby halo stars. 

\subsection{Tides and dynamical friction\label{sec:s21}}

Two effects are important for accreted satellites to reach the inner halo and the disk. The first is dynamical friction. As the satellite moves through the fluid of lighter particles of the host halo, they get pulled towards the satellite, which in the meantime moves along. The resulting overdensity or ``gravitational wake'' behind the satellite leads to a steady deceleration of the latter. The deceleration of a satellite of (point) mass $M_\text{s}$ moving at velocity $\bm{v}_s$ on a straight orbit through an infinite, homogeneous and isotropic field of particles with mass density $\rho_\text{host}$ can be expressed through Chandrasekhar's analytic formula \citep{cha43}:
\begin{equation}\label{eq:e1}
	\frac{d\bm{v}_\text{s}}{dt}=-4\pi G^2M_\text{s}\ln\Lambda\,\rho_\text{host}(<v_\text{s}) \frac{\bm{v}_\text{s}}{\vert\bm{v}_\text{s}\vert^3}.
\end{equation}
Here, $G$ is the gravitational constant, $\rho_\text{host}(<v_\text{s})$ the density of field particles at the position of the satellite with velocities less than $\vert\bm{v}_\text{s}\vert$ and $\ln\Lambda$ is the Coulomb logarithm, named so in analogy with the equivalent logarithm in the field of plasma physics. Note that the drag force, $M_\text{s}\frac{d\bm{v}_\text{s}}{dt}$, is proportional to $M_\text{s}^2$, so that massive satellites loose their energy and angular momentum much faster than low massive ones. Equation~\ref{eq:e1} further predicts the merging timescale to depend on the masses of the satellite and host halo as \citep[e.g.][eq.~8.13]{bin08}
\begin{equation}\label{eq:e2}
	\tau_\text{merge}\propto\Bigl(\ln\Lambda\frac{M_\text{s}}{M_\text{host}}\Bigr)^{-1}.
\end{equation} 
Besides assuming linear motion of a point mass through an uniform and isotropic background of particles, Chandrasekhar's formula also neglects any gravitational attraction of these particles for one another. Therefore, in general, equations~\eqref{eq:e1} and~\eqref{eq:e2} can not describe accurately the orbital decay of extended, mass-loosing satellites moving on eccentric orbits within a live host halo in $N$-body simulations. For example, while \citet{sal07} found that the infall timescales of satellites onto host halos in their $N$-body/gasdynamical simulations follow the trend with mass ratio given by equation~\eqref{eq:e2}, \citet{boy08} found a non-linear dependence in their $N$-body simulations scaling as $(M_\text{s}/M_\text{host})^{-1.3}$ (i.e. the satellite sinking more slowly than predicted by the standard formula). Another uncertainty is the choice of the Coulomb logarithm.
The term $\ln\Lambda$ is given by
\begin{equation}\label{eq:e3}
	\ln\Lambda=\ln\,\Bigl(\frac{b_\text{max}}{b_\text{min}}\Bigr),
\end{equation}
where $b_\text{min}$ and $b_\text{max}$ denote the maximum and minimum impact parameters, i.e. the maximum and minimum distances of closest approach between the satellite galaxy and the field stars of the host galaxy. Usually, $b_\text{min}$ is set to the size of the satellite galaxy \citep[i.e. its virial radius,][]{whi76}, while the choice of $b_\text{max}$ is less clear and somewhat arbitrary \citep[][\S~8.1]{bin08}. \citet{has03} argued that $b_\text{max}$ should be varied according to the actual separation of the satellite from the host galaxy's center in order to account for the density gradient of the host halo in $N$-body simulations. However, \citet{fuj06} showed that this approach predicts too fast sinking times of satellites which suffer significant mass loss, because recently stripped particles would contribute to the drag on the satellite. 

\begin{figure*}[ht]
	\includegraphics[width=0.95\textwidth]{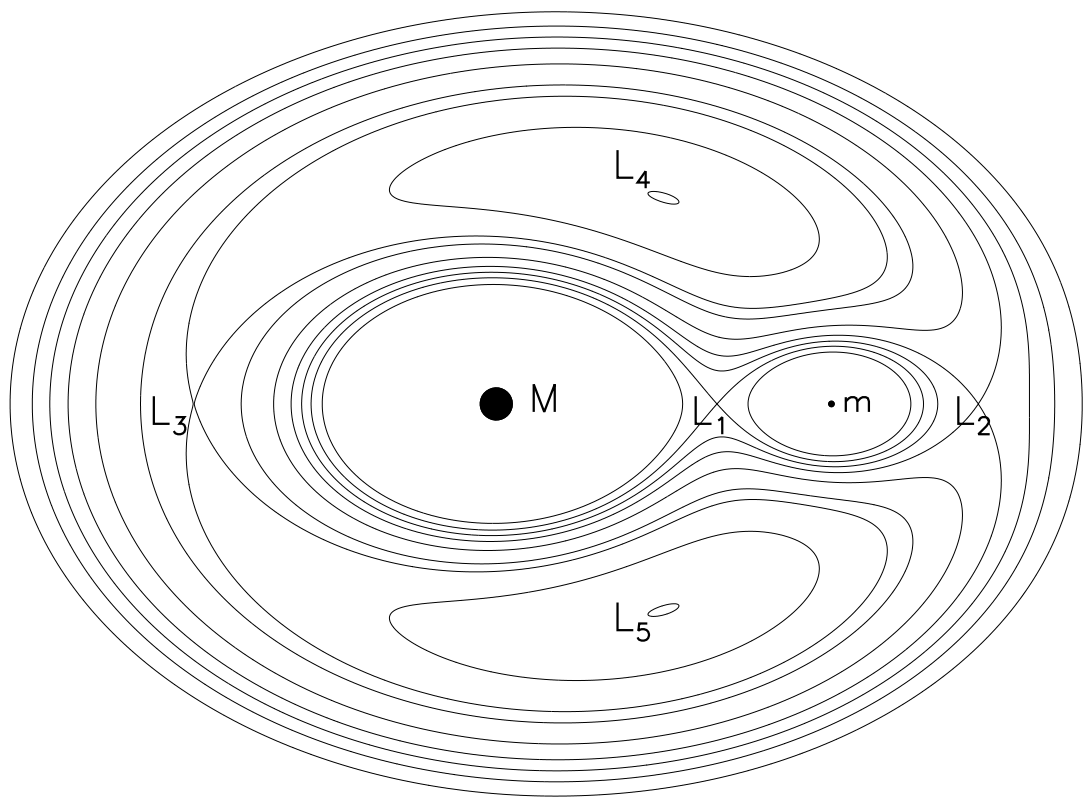}
	\caption{Contours of constant effective potential in the restricted three-body problem (Jacobi-Hill theory) for satellite-to-host halo mass ratios $m/M=1/5$. Note the non-circular form of the Roche surface, which contains L$_1$, and the asymmetry of L$_1$ and L$_2$ with respect to $m$. Stars with $E_J<\Phi_\text{eff}(\text{L}_1)$ are bound to the satellite.}
	\label{fig:f1}
\end{figure*}

Mass loss of the satellite is the second important effect and occurs primarily at each pericentric passage through tidal shocks, but also at other parts of the orbit due to the limitation of bound satellite particles to stay within a tidal surface. If the satellite is on a circular orbit, Jacobi-Hill theory predicts that stars with Jacobi energy $E_J=\frac{1}{2}v^2+\Phi_\text{eff}$ stay bound to the satellite, if they lie inside the last closed zero velocity surface, the so-called Roche surface, at which $E_J$ equals the effective potential \citep[Figure~\ref{fig:f1} and][\S~8.3]{bin08}. The latter is given through $\Phi_\text{eff}(\bm{r})=\Phi(\bm{r})-\frac{1}{2}\vert\omega\times\bm{r}\vert^2$, where $\Phi({r})$ is the gravitational potential and the second term describes the centrifugal force with the angular speed $\omega$ of each mass around the common center of mass. The Roche surface includes the two Lagrange points L$_1$ and L$_2$, which are saddle points of the effective potential and are inversion symmetric with respect to the satellite's center only for very small satellite masses \citep[][]{choi07}. Stars bound to the satellite and having Jacobi energy $E_J$ can escape the satellite, if weak perturbations elevate $E_J$ above $\Phi_\text{eff}(\text{L}_1)$ or $\Phi_\text{eff}(\text{L}_2)$. So, mass loss occurs near the two Lagrange points, leading to the formation of a leading and trailing tidal tail. Such tails can be observed as a spectacular example stretching from the Sagittarius dwarf spheroidal \citep{iba01,maj03,bel06a} or the globular cluster Palomar 5 \citep{ode01,grill06}. The leading tail contains stars at lower energies than the trailing tail, so each time stars get stripped from the satellite, e.g. through tidal shocks at pericenter, roughly two groupings of stars at distinct energies result \citep{joh98,meza05,choi07}. \citet{war08} proposed that this energy difference could be used for the distinction between leading and trailing stars even after significant dispersion throughout the Galaxy. As pointed out by \citet{choi07}, (dark matter) particles in the leading tail continue to be decelerated by their satellite's gravity and fall towards the central regions of the host. It has not been studied yet, however, if this could provide another mechanism for stars to populate the inner halo, even if their satellite stays hung up in the outer halo. 

\subsection{The contribution to the inner halo\label{sec:s22}} 
Both the adiabatic growth of the host halo and the energy loss of the satellite through dynamical friction have the effect of its orbit spiraling in towards the Galactic center. There is still debate whether the orbits also circularize during infall, and if so, whether dynamical friction causes this circularization \citep[e.g.][]{col99,gil04}. However, the decay of the satellite's orbit leads to a stronger tidal field and increases the mass loss, which in turn decreases dynamical friction. Using a fully analytical model, \citet{zhao04} estimated this interaction between dynamical friction and mass loss to delay the merging of typical present day satellites ($10^7-10^9M_\odot$) with initial distances $d>20$ kpc to more than a Hubble time. In support of this, \citet{col99} already argued that the effect of mass loss outperforms dynamical friction and makes it unlikely that remnants of satellites with $M_\text{s}/M_\text{host}>0.1$ have reached the inner halo yet.

Numerical simulations of the hierarchical buildup of Milky Way-like galaxies in the $\Lambda$CDM context seem to support these dynamical friction arguments. A consistent finding is that the inner halo consists of particles stripped from a few significant contributors. These are more massive and have been accreted much earlier ($\gtrsim$8 Gyr ago) than the satellites that remain today \citep{bul05,font06,sal07,kaz08,read08}, although the possibility exists that disruption of the massive satellites is not yet completed and remnants still exist in the inner halo \citep{coo10}. Stated otherwise, ``\textit{the `building blocks' of the stellar halo were on average more massive and were accreted and disrupted much earlier than the population of satellites that survive until the present.}'' \citep{sal07}. So, in a cosmological context, the presently observed satellites are not assumed to be representative for the ones already accreted into the inner halo. Therefore it is not surprising that present-day satellites exhibit different chemical compositions than halo stars. The latter tend to have higher [$\alpha$/Fe] ratios at a given [Fe/H] \citep{venn04}, where $\alpha$ stands for the average abundance of the elements Mg, Si, Ca and Ti. This points towards rapid star formation in their progenitor systems, because the $\alpha$ elements are mainly produced in type II supernova (SN) explosions on a short timescale ($\sim10^7$ yr), while only after a longer timescale of $\sim10^9$ yr SNeIa start to lower [$\alpha$/Fe] again through expelling additional Fe into the ISM. Recently, however, \citet{nis10} reported a dichotomy of nearby halo stars ($\langle d\rangle=115$ pc) with respect to [$\alpha$/Fe], with the high-$\alpha$ stars being more gravitationally bound than the low-$\alpha$ ones. This could indicate that low mass dwarf galaxies have contributed to the inner halo, although to what extend is still to be determined. A possible mechanism for dwarf galaxies to reach the inner halo is infall in massive, loosely bound groups, as pointed out by \citet{li08}. According to \citet{read08}, this allows even low mass satellites to merge into the inner halo in less than a Hubble time. \citeauthor{read08} also have shown that -- assuming isotropic infall -- a typical Milky Way halo will have one out of three mergers at a given mass within 20$^\circ$ inclination to the disk plane. While the high-inclination satellites would primarily contribute to the Milky Way stellar halo \citep{bul05}, the low-inclination ones would be dragged completely into the disk plane by dynamical friction \citep[Fig.~6 in][]{read08}, which results in a richly structured accreted thick disk \citep[see also][]{abadi03b,meza05}. In all cases the encounters would heat the pre-existing thin disk\footnote{Thereby only very massive high-redshift satellites would be able to form a thick disk from a thin disk through heating \citep{kaz08}.} and produce several morphological features like bars, flares and warps \citep{abadi03b,read08,vil09}, or even overdensities of stars at distinct rotational velocities that travel as waves through velocity space \citep{min09}. 

It is interesting to ask how one can distinguish between \textit{accreted} and \textit{in situ} thick disk stars in a metal-poor local sample. This would give us clues about the fraction of accreted stars versus those that were born within the Milky Way's potential well and contributed to the thick disk through other processes. Mainly three such processes have been discussed in the literature: (i) heating of a pre-existing thin disk \citep[e.g.][]{vil08,kaz08}, (ii) radial mixing, i.e. migration of stars both from inner and outer radii through the solar neighborhood \citep{sel02,hay08,sch09}, or (iii) \textit{in situ} star formation during a gas rich merger \citep{brook04}. By directly comparing the orbital properties of stars from four simulations of these thick disk formation scenarios, \citet{sal09} found clear differences in the eccentricity distribution between the generated thick disks. While accreted stars always contribute to the high-eccentricity tail of the distribution, \textit{in situ} stars have low eccentricities $e\sim0.2-0.3$, regardless which of the mechanisms (i) to (iii) gave them thick disk properties. Given that a certain fraction of stars in the Solar neighborhood has highly eccentric orbits \citep{chi00} and assuming that they have indeed been accreted, we can further ask if we can trace back any of theses stars to a common origin. In other words, do any of these stars belong to the same stellar streams? Assuming a halo solely built through accretion, \citet{helm99b} predicted the number of individual streams passing the solar neighborhood to be
\begin{equation}\label{eq:e4}
	N_\text{streams}\sim[300-500](t/10\text{Gyr})^3,
\end{equation}
where $t$ is the time since disruption. In reference to the dynamical friction arguments made above, these streams are expected to stem mainly from a few progenitors only. The Milky Way seems to have a quiet merger history since $t\sim10-12$ Gyr (corresponding to the age of the thick disk). In such a time any debris from an accreted satellite is expected to be spread over several tens of kpc. Therefore one cannot expect to find many of the halo streams predicted by equation~\eqref{eq:e4} as spatially coherent structures in the solar neighborhood \citep{sea08}, although hints for a spatial clumping of one kinematically identified substructure have been found \citep[][]{kle09}. Instead, other search strategies for stellar streams have to be applied, which I describe in the following section.

\section{Search strategies for halo streams\label{sec:s3}}
\subsection{Integrals of Motion and Action-Angle Variables\label{sec:s31}}
The \textit{collisionless Boltzmann equation}
\begin{equation}\label{eq:e5}
	\frac{d}{dt}f(\bm{r},\bm{v},t)=\frac{\partial f}{\partial t}+\sum_{i=1}^3\Bigl(v_i\frac{\partial f}{\partial x_i}-\frac{\partial\Phi}{\partial x_i}\frac{\partial f}{\partial v_i}\Bigr)=0
\end{equation}
is the fundamental equation of stellar dynamics. It assumes that encounters between stars can be ignored so that every star moves under a mean gravitational potential $\Phi(\bm{r},t)$ (therefore the term ``collisionless''), where $\bm{r}=(x_1,x_2,x_3)$, and $\bm{v}=(v_1,v_2,v_3)$ denote the position and velocity vector of a star with respect to the center. 
$f(\bm{r},\bm{v},t)$ is the distribution function describing the density of stars in phase-space similar to the distribution of atoms in a gas. Equation~\eqref{eq:e5}
states that the phase-space density $f$ around the phase-space point of any particular star is constant. It follows that, as a sample of stars which is initially concentrated into a small phase-space volume (like in a satellite galaxy) disperses in configuration space, at any particular point in phase-space they will grow together kinematically in an infinitesimal volume around that point. For example, \citet{helm99b} have shown that the velocity dispersions at each point along the track of a \textit{single} stream of stars lost at the same passage decrease overall with $t^{-1}$ with some periodic oscillations due to the spatial density enhancements at the turning points of the orbit. An observer, however, will always measure the coarse-grained phase-space density $\langle f\rangle$, which is the average phase-space density in his finite observational volume. In the inner halo, where orbital timescales are short, $\langle f\rangle$ includes the contribution from \textit{multiple} streams of a single progenitor.
In a given local volume, we eventually will find stars belonging to each of these streams, while an infinitesimal volume only contains stars belonging to one individual stream. This is known as phase mixing. It is caused by the fact that for any individual stream, the phase-space regions with initially high density stretch out and get thinner; the coarse-grained density decreases with time until a final equilibrium state is achieved \citep[e.g.~Figure~1 in][]{tre99}. Phase mixing means in practice that in a given volume the stellar density of tidal debris from a satellite decreases and its velocity dispersion increases with time, so that individual streams can no longer be separated in velocity space. The broadening in the velocity distribution is caused by the initial differences in the orbital phases between stars stripped at different epochs. In the disk for example the broadening in the velocity distribution is most prominent in the vertical $W$-component, because the vertical frequency with respect to the disk is shorter than the horizontal frequencies. For the angular $V$-component, \citet{helm99b} calculated that complete phase mixing would give rise to an observed velocity dispersion of 50--150 km s$^{-1}$, depending in proportion on the initial velocity dispersion of the satellite. 

\begin{figure*}[ht]
	\centering
		\includegraphics[width=0.95\textwidth]{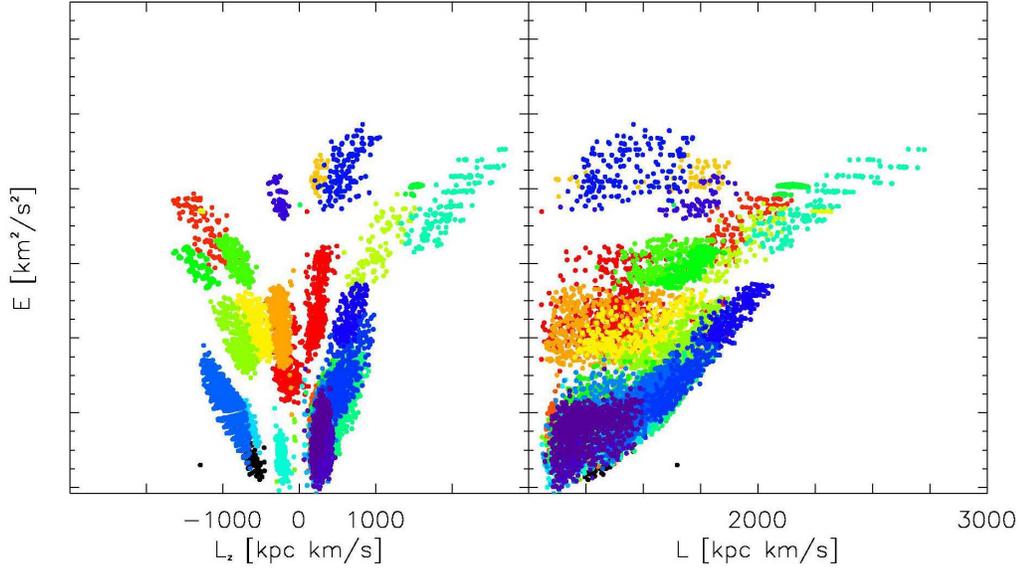}
	\caption{Simulated $(E,\vert\bm{L}\vert,L_z)$-distribution of stars that have been stripped from 33 different (color-coded) satellites and observed 12 Gyr within 2 kpc from the sun. Simulated errors as expected for the \textit{Gaia} mission have been added to the data. Same data as in \citet[][Fig.~4]{helm00}, except for the smaller size of the observational volume.}
	\label{fig:f2}
\end{figure*}

The problem of phase mixing can be avoided, if one switches to integrals-of-motion-space (IoM-space). An \textit{integral of motion} $I(\bm{r},\bm{v})$ is a function of the phase-space coordinates that is constant along any stellar trajectory in phase-space. It does not depend on time:
\begin{equation}\label{eq:e6}
	I(\bm{r}(t),\bm{v}(t))=\text{const}.
\end{equation}
An integral of motion is called isolating if it isolates the points on a star's trajectory from neighboring points in phase space, contrary to non-isolating integrals which are infinitely many-valued \citep[for more details see e.g.][]{oll62,bin08}. Classical isolating integrals of motion include the energy $E$ in every static potential, the three components of the total angular momentum in a spherically symmetric potential, or one of its components, e.g.~$L_z$, in an axisymmetric potential. According to \textit{Jeans theorem}, in a steady state any function of $n$ integrals of motion automatically solves the collisionless Boltzmann equation \eqref{eq:e5}, because
\begin{equation}\label{eq:e7}
 \frac{d}{dt}f\bigl(I_1(\bm{r},\bm{v}),...,I_n(\bm{r},\bm{v})\bigr)=\sum_{i=1}^n\frac{\partial f}{\partial I_i}\frac{dI_i}{dt}=0.
\end{equation}
The coarse-grained phase-space density -- written as a function of the integrals of motion -- evolves asymptotically to a state that satisfies Jean's theorem. Therefore, a subpopulation of stars initially confined to a small phase-space volume will remain clumped in the IoM-space (Figure~\ref{fig:f2}).

Additional insights can be gained from studying the dispersion of stream particles in action-angle coordinates $(\bm J,\bm\Theta)$ \citep{tre99, helm99b, bin08}. In Hamiltonian mechanics, actions and angles form a special set of canonical variables that lead to a simple form of the canonical equations of motion:
\begin{equation}\label{eq:e8}
  \frac{d\;\bm \Theta}{d\;t}=\frac{\partial H(\bm J)}{\partial\bm J}\equiv\bm\Omega(\bm J)    \qquad   \frac{d\;\bm J}{d\;t}=-\frac{\partial H(\bm J)}{\partial\bm\Theta}=0,
\end{equation}
where the components of $\bm\Omega(\bm J)$ are the orbital frequencies.
Neglecting the influence of the satellite's potential on the individual stars, the actions $\bm J$ of a star remain constant, while the angles increase linearly with time.
The cloud of escaped stream particles therefore spreads in only three of the six phase space dimensions at rates which can be shown to depend on the eigenvalues of the Hessian matrix of the Hamiltonian\footnote{The elements of the Hessian matrix are given by $H_{ij}\equiv\frac{\partial^2H}{\partial J_i\partial J_j}$.} and the initial spread in the actions. For St{\"a}ckel potentials that resemble disk-like flattened axisymmetric potentials, it can be shown analytically that the actions are functions of the three isolating integrals $E$, $L_z$ and $I_3$. Again it follows that the stream stars remain clumped in the space of these integrals, even if the potential underwent slow changes in the past. The reason is that usually two or three action variables are adiabatically invariant (like $J_\Phi=L_z$ in an axisymmetric potential). 

\begin{figure}[ht]\sidecaption
	\centering
		\includegraphics[width=0.45\textwidth]{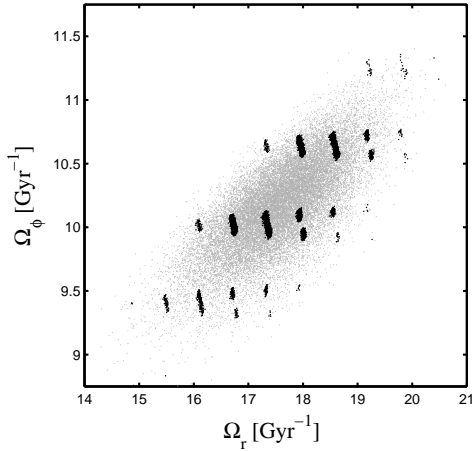}
	\caption{$\bm\Omega$-space distribution of stream particles lost from a satellite without modeled self-gravity on an orbit in a spherical static potential. Grey dots show the distribution of all satellite particles after 10 Gyr, whereas the black dots only show the particles inside a sphere of 4 kpc radius. Note the alignment of the patterns along lines of constant frequencies $\Omega_r$ and $\Omega_\Phi$. Adopted from \citet[][Fig.~5]{gom10a}.}
	\label{fig:f3}
\end{figure}

Attempts to use the action-angle variables directly to search for fossil remnants of past accretion events have until now only been made sparsely \citep[e.g. by][]{chi00}. The reason might be that action-angle variables are calculated readily only for integrable, time-independent potentials, while for real stars in the unknown Galactic potential they have to be constructed via sophisticated numerical methods \citep[e.g.][]{bin84,mcg90}. Theoretically, the space of orbital frequencies seems to be a promising one \citep{mcm08,gom10a}. In a static potential, stars stripped at time $t=0$ and observed at time $t$ can be identified as distinct clumps -- each representing an individual stream and separated by regular distances $\delta\Omega_i$ that directly measure $t$ through $\delta\Omega_i=\frac{2\pi}{t}$ (Figure~\ref{fig:f3}). The reason is that any observational volume, such as the solar neighborhood, only picks out those stars that have moved a certain amount in an angular direction, e.g. azimuth $\Phi$, plus or minus an integer number of completed angular periods, like rotations about the Galactic center. The number of these integers is given by $\Omega_it/2\pi$, so the number of patches in $\bm\Omega$-space\footnote{Usually, the vertical frequency $\Omega_z$ is omitted because $\Omega_z\gg\Omega_R\; \vee\;\Omega_z\gg\Omega_\Phi$, implying that the number of streams in the $z$-direction is much larger, and so after some time, much harder to see unless one has very high numerical resolution (or the orbit is elongated in the $z$-direction).} increases proportional with $t^2$. Meanwhile, they get smaller proportional to $t^{-2}$, because the width in one dimension is given by $\Delta\Theta_i/t$, where $\Delta\Theta_i$ is the range in angle component $i$ which in turn is determined by the size of the observational volume \citep{mcm08}. \citet{gom10a} have shown that also for a time-evolving host potential and even a N-body simulation of the accretion of a satellite onto a live disk, the clumps in frequency space are conserved, although their regularity breaks down in the latter case. In addition, the timescale of growth of the host potential seems to leave its imprint as a curvature in the alignment of the patches -- a clear advantage over the classical integrals $E$ and $L_z$, that according to \citet{pen06} can only constrain the present Galactic potential. Recently, \citet{gom10b} applied the stream search in frequency space to a mock \textit{Gaia} catalog that contained tidal debris from 42 disrupted satellites. In their simulations, these authors included observational errors expected for \textit{Gaia} as well as a time-varying potential of the halo, bulge and disk. \citeauthor{gom10b} found that only for small observational errors (relative parallax errors less than 0.02), a discrimination between adjacent streams in frequency space and a clear determination of the time since accretion are possible. In their simulations, they were able to compute accretion times for $\sim30\%$ of the identified satellites, accurate to 15--25\%. The estimated times are always lower limits and the difference to the actual time since accretion depend on the form of the gravitational potential and its evolution. 

The actions $\bm J$, besides having the advantage of being constant and adiabatically invariant once the influence of the satellite's gravitation has ceased, also allow the identification of individual streams as different patches \citep{mcm08}. However, the patches are not aligned along lines of constant $J_i$ like in the case of the $\Omega_i$s (in a static potential), because the $\Omega_i$s depend on all three $J_i$s, so that the constraints on $\bm\Omega$ also affect $J_z$. Therefore, \citet{gom10a} found that individual streams are less clearly separable in $(J_r,J_\Phi)$-space, comparable to the space of classical integrals $(E,L_z)$. However, the large expected number of streams passing the solar neighborhood given by equation~\eqref{eq:e4}
could make it difficult to detect individual streams from a given satellite, especially when observational errors are considered. Therefore, in practice, it might be better to use the calssical integrals of motion $(E,\vert\bm L\vert,L_z)$ to first detect tidal debris, because individual streams of a satellite roughly contribute to the same clump in IoM-space, raising the chances of identifying the individual satellites. Afterwards, $\bm\Omega$-space could be used to compute the time since accretion \citep{gom10b}.

Indeed, the classical integrals of motion have been used successfully for finding tidal debris in the solar neighborhood \citep{helm99a,chi00,refi05,kep07,smi09}; however, some limitations have to be considered \citep[pointed out by][]{helm06}: First, the total angular momentum $\vert\bm L\vert$ is only strictly conserved in a spherical potential which is not the case for the Milky Way. However, this seems to be a minor point, because even for stars that move in axisymmetric flattened potentials, $\vert\bm L\vert$ is approximately conserved and similar to the third integral $I_3$, for which no analytic expression exists \citep[][\S~3.2.2]{bin08}. Also, \citet{chi00} showed that the distribution of halo stars in the space spanned by isolating integrals of motion in an aspherical St{\"a}ckel-type potential can be closely mapped into the IoM-space of a spherical potential. A second concern is that continuing accretion and merging events would have changed the potential steadily and altered the energy of an orbiting satellite. This emphazises the advantages of adiabatic invariants such as the actions (among them $J_\Phi=L_z$). However, the clumpiness of the local IoM-space \citep{mor09,kle09} suggests that (i) violent relaxation might have had a less smoothing effect on the signatures of the inner halo stars than expected and/or (ii) only a couple of early major mergers have contributed to the inner halo, in agreement with predictions from $\Lambda$CDM simulations and a direct effect of dynamical friction acting primarily on the most massive satellites \citep[e.g.][and \S~\ref{sec:s22} in this review]{helm03}. Both Dynamical friction and the gravitational influence of the satellite also lead to the problem that stars lost at different passages end up having different energy levels, in this way creating multiple signatures in the $(E,L_z)$-space \citep{meza05,choi07}. Finally, to calculate the energy, one needs the full 6D phase space information $(\bm{r},\bm{v})$, which can be observed, plus the potential $\Phi(\bm{r})$, which is desired, but not directly observable.

\begin{figure*}[ht]
	\centering
		\includegraphics[width=0.95\textwidth]{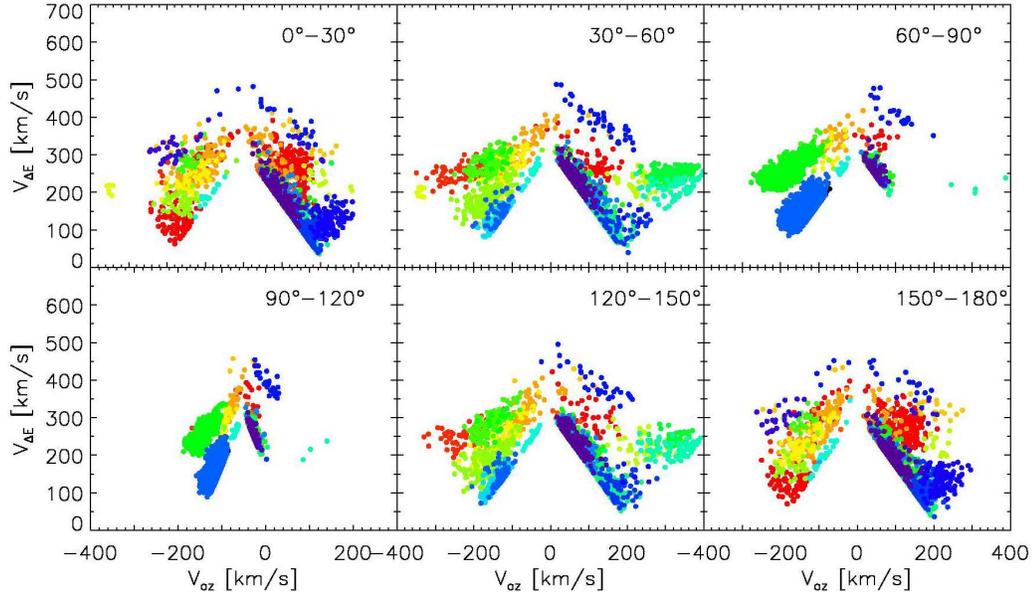}
	\caption{$(\nu,V_\text{az},V_{\Delta\text{E}})$-distribution of the same stars as in Fig.~\ref{fig:f2}. The same color coding has been used. The stars have been binned into $30^\circ$ wide $\nu$-slices, as indicated in each panel, and displayed in $(V_\text{az},V_{\Delta\text{E}})$. Note that these quantities are derived in a spherical potential, although the true potential in this case includes a disk \citep[][Model~I]{helm00}. Also note the symmetry of multiple streams from the same satellites with respect to $\nu=90^\circ$, which follows from the symmetry in the $(V,W)$-distribution of phase-mixed tidal debris (Fig.~\ref{fig:f5} below) and equation~\eqref{eq:e9a}.}
	\label{fig:f4}
\end{figure*}

Therefore alternative ``effective'' integrals of motion have been proposed \citep{helm06,ari06,det07,kle08,kle09}. The idea behind such approaches is to find parameters that characterize stellar orbits and can be easily expressed through the observables. For example, \citet{helm06} proposed to look for stellar streams in a space spanned by the apocenter, pericenter and angular momentum $L_z$ (the APL-space). Moving groups then cluster around lines of constant eccentricity. However, to calculate the eccentricity, a guess of the true potential is required. \citet{ari06}, \citet{det07}, \citet{kle08} and \citet{kle09} followed a similar approach in the sense that they tried to find stars with the same orbital eccentricity. Their strategy in finding nearby stellar streams in velocity space is based on the Keplerian approximation for orbits developed by \citet{dek76}; in this approximation, the potential $\Phi(R)$ is expanded with respect to $\frac{1}{R}$, and the effective potential, $\Phi_\text{eff}(\vert\bm L\vert,R)\equiv\Phi(R)+\vert\bm L\vert^2/(2R^2)$, takes on a form similar to the one in the Kepler problem. The asumption is that stars in the same stellar stream move on orbits that stay close together, which is justified by numerical simulations of satellite disruption \citep{helm06}. Further assuming a spherical potential and a flat rotation curve, these stars should form a clump in the projection of velocity space spanned by
\begin{align}
\label{eq:e9a}\nu&\equiv\arctan\bigl(\frac{V+V_\text{LSR}}{W}\bigr),\\
\label{eq:e9b}V_\text{az}&\equiv\begin{cases}~~\sqrt{(V+V_\text{LSR})^2+W^2} \qquad &\nu\leq180^\circ\\
-\sqrt{(V+V_\text{LSR})^2+W^2} \qquad &\nu>180^\circ, \end{cases}\\
\label{eq:e9c}V_{\Delta\text{E}}&\equiv\begin{cases}\sqrt{U^2+2(V_\text{LSR}-V_\text{az})^2}\qquad &\nu\leq180^\circ\\
       \sqrt{U^2+2(V_\text{LSR}+V_\text{az})^2}\qquad &\nu\leq180^\circ.\end{cases}\end{align}
$\nu$ is the angle between the orbital plane (which is fixed in a spherical potential) and the direction towards the North Galactic Pole and ranges from $0^\circ-180^\circ$. Stars with $\nu>180^\circ$ are treated as stars moving on retrograde orbits in a plane with inclination angle $\nu-180^\circ$. $V_\text{az}$ is related to the angular momentum $\vert\bm L\vert$ and defines the guiding center radii of the stars, $R_0=R_\odot\frac{V_\text{az}}{V_\text{LSR}}$, in the plane labeled by $\nu$. $V_{\Delta\text{E}}$ is a measure of a star's eccentricity $e$ through
\begin{equation}\label{eq:e10}
e=\frac{1}{\sqrt{2}V_\text{LSR}}V_{\Delta\text{E}}
\end{equation}
 \citep[eq.~11 in][]{kle09}. Also, it can be shown that $V_{\Delta\text{E}}^2$ is adiabatically invariant, because it approximately relates to the radial action integral $J_R$ \citep[eq.~13 in][]{kle09}\footnote{In the formalism of \citet{ari06} and \citet{kle08}, slightly different notations have been used, which follow easily under the assumption of motions in the orbital plane $z=0$ or, equivalently, $\nu=90^\circ$ only. Note also that in the works of \citet{det07} and \citet{kle09} only the upper definition of equation~\eqref{eq:e9c} has been used, which still allows identification of substructure, but not the calculation of eccentricities from equation~\eqref{eq:e10}.}. Equations~\eqref{eq:e9a}--\eqref{eq:e9c} fully define an orbit in the Keplerian approximation, because its turning points are given by 
\begin{equation}\label{eq:e11}
\frac{R_\text{min}}{R_0}=\frac{1}{1+e}\qquad\text{and}\qquad \frac{R_\text{max}}{R_0}=\frac{1}{1-e}.
\end{equation}
\citet{dek76} has shown that her approximation holds for eccentricities up to $e=0.5$ ($V_{\Delta\text{E}}^2\approx150$ km s$^{-1}$), although in practice, the clumping in the space spanned by the quantities \eqref{eq:e9a}--\eqref{eq:e9c} seems to remain even for stronger deviations from circularity (Figure~\ref{fig:f4}).

\subsection{The use of Velocities and Positions\label{sec:s32}}
\begin{figure*}[ht]
	\includegraphics[width=0.95\textwidth]{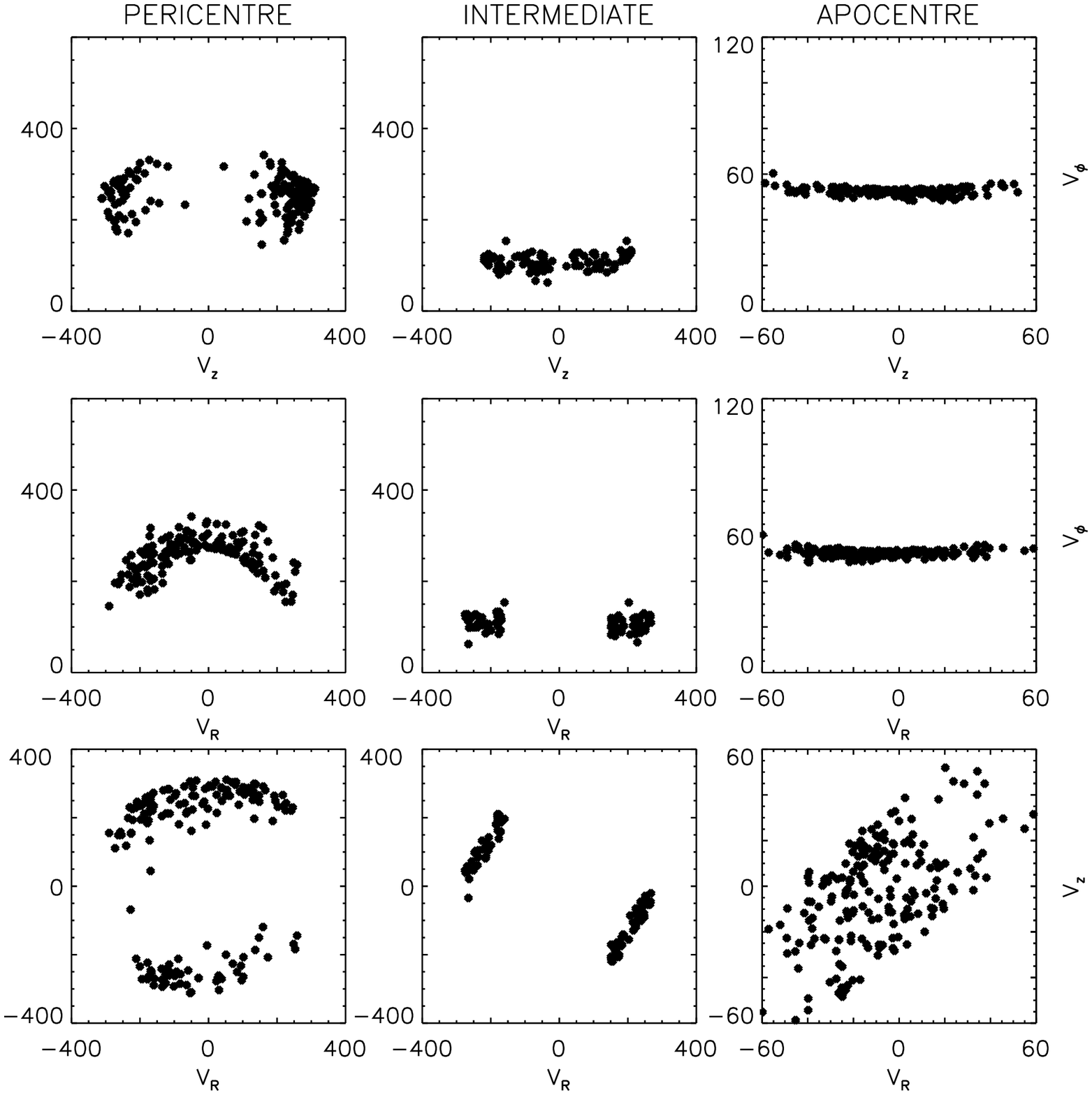}
	\caption{Simulated velocity space distribution of satellite stars 13.5 Gyr after becoming unbound to their progenitor system. The figure only plots stars within a 3 kpc wide box centered on the progenitor's pericenter ($r=6.0$ kpc, \textit{left panel}), apocenter ($r=37.0$ kpc, \textit{right pane}l) or within (\textit{middle panel}). Note the banana shape of stars near their apo- and pericenters. Note that in the solar neighborhood $v_R\approx U$, $v_\phi\approx V$ and $v_z=W$. Adopted from \citet[][Fig.~5]{helm99b}. }
	\label{fig:f5}
\end{figure*}

As noted above, stream stars passing the solar neighborhood have very similar azimuthal frequencies or, equivalently, angular momenta. Therefore these stars show a narrow distribution of azimuthal velocities. In this respect they behave exactly like the classical moving groups in the solar neighborhood, that emerge either from dissolved open clusters or through dynamical resonances with the non-axisymmetric potential. However, moving halo groups often have a distinct banana-shaped $(U,V)$ velocity distribution, which is a consequence of their eccentric orbits \citep{helm99b,helm06,vil09}: The slightly different orbital phases of the stars in the observed volume can result in some stars being at their apocenters ($U=0$), while the others either move
away from ($U>0$) or towards ($U<0$) it (Figure~\ref{fig:f5}). For the same reasons, if the pericenter of some stars lies in the observational volume, a similar, but switched and stronger curved banana pattern results (because stars move fastest near their pericenters). Compared to \textit{in situ} thick disk stars, accreted stars mostly occupy the outskirts of the $(U,W)$-distribution \citep{vil09}. 

An even clearer separation between tidal debris and metal-poor disk stars can be obtained by observing stars out to large cylindrical radii $R$. Assuming a constant rotation curve implies $L_z\propto R$ for disk stars, whereas stream stars display a nearly constant angular momentum that depends mainly on the initial conditions of their progenitor's orbit (like the orbital inclination), but not on $R$. Therefore, by plotting $R$ versus $L_z$, the disk stars discriminate themselves from the stream stars by their trend of rising $L_z$ as $R$ increases \citep[Fig.~6 in][]{vil09}. 

Finally, heliocentric line-of-sight velocities $v_\text{los}$, which are much easier to obtain than full space motions, can be used to detect stellar streams against the disk star background. In this case, circular rotation of the disk stars implies a sinusoidal dependence of $v_\text{los}$ on galactic longitude $l$ \footnote{$v_\text{los}=(U-U_\odot)\cos b\cos l+(V-V_\odot)\cos b\sin l+(W-W_\odot)\sin b$; for circular low latitude orbits this implies $v_\text{los}\sim(V-V_\odot)\cos b\sin l$.}, while again the stream stars dominate the wings of the $v_\text{los}$ distribution and should become apparent after subtraction of the mean disk rotation \citep{vil09}.

\section{Moving halo groups\label{sec:s4}}
\subsection{Current state of research\label{sec:s41}}
\subsubsection{The early years\label{sec:s411}}
Since \citet{pro1869} it has been known that among the nearby stars some groups exist that share the same streaming motion, therefore called stellar streams or moving/kinematic groups. It was hypothezised early that such moving groups result from the breakup or dispersion of open clusters \citep[see e.g.][and references therein]{egg58}, an effect that also would apply to halo globular clusters \citep{oort65}, satellite galaxies or other accreted building blocks of the Galaxy like the \citeauthor{searle78} fragments. In 1959, \citeauthor{egg59} discovered a group of five subdwarfs within 20 pc -- among them the star RR Lyrae -- with large space motions moving towards a common convergence point. The group was named the Groombridge 1830 group after the star Groombridge 1830 (HR 4550), which lags the Local Standard of Rest with $V\approx-130$ km s$^{-1}$. In the years that followed, Eggen extended his observations of subdwarfs in number and distance and established the existence of high-velocity, metal-poor moving groups \citep[e.g.][and references therein]{egg77}. Two of these, the groups named after Kapteyn's star and Arcturus, have only recently been investigated in great detail (see \S~\ref{sec:s413}). In the early 90s, despite some false detections of halo substructure, the evidence for nearby kinematic halo streams interpreted as tidal debris increased \citep[for a review see][Sect.~5.1]{maj93}. \citet{tre93} computed the fractional volume containing one or more tidal streams to be $\approx1$ at the Solar radius, if the halo would have been built solely through merging and accretion. 10 years later, \citet{gou03} should derive from statistical arguments on a set of 4588 halo stars that there are about $\sim400$ cold streams present in the solar neighborhood, whereby no stream could contain more than 5\% of the local stars. This number is fully consistent with the predictions of equation~\eqref{eq:e4} and has until now not been proved false. More possible streams were found when \citet{pov92} analyzed the clustering of 206 halo stars in $(E,L_z,\text{[Fe/H]})$-space. These authors justified their approach with the argument that ``...because of the high incidence of chaotic orbits among halo stars...the usual criterium of identifying group members by the similarity of their $V$-velocities breaks down.''

\subsubsection{The Helmi stream\label{sec:s412}}
In 1999, \citet[][hereafter H99]{helm99a} published their seminal paper on the discovery of two single streams stemming from the same progenitor, that ``...had a highly inclined orbit about the Milky Way at a maximum distance of $\sim16$ kpc, and it probably resembled the Fornax and Sagittarius dwarf spheroidal galaxies.'' The two streams are clearly separated in the $(U,V)$-plot, but occupy the same region in the $(L_z,L_\perp)$-plane ($\sim(1200,2200)$ kpc km s$^{-1}$.) The stream stars are on prograde eccentric orbits ($R_\text{peri}\sim7$ kpc, $R_\text{apo}\sim16$ kpc) that reach distances up to $\vert z\vert_\text{max}\sim13$ kpc above or below the plane. Originally, \citetalias{helm99a} concluded that about 10\% of the local halo stars would belong to this stellar stream. However, this number was questioned when \citet{chi00} analyzed 728 stars with full space motions (accurate to $\sim20\%$) and [Fe/H]<-1 which were drawn from the \citet{bee00} sample of 2106 metal-poor stars without kinematic bias. Although this meant a sample size three times as large as H99's, they found practically the same number of stream members. The \citeauthor{bee00} catalog was re-examined by \citet{refi05}, but now concentrating on the classical search strategy for moving groups, namely picking out stars with strong correlations in velocity space. Among the 5\% of the fastest moving halo stars, \citeauthor{refi05} found a kinematic group consisting of 3 stars with similar angular momenta, but somewhat lower binding energy than the H99 streams. They hypothezised that their moving group could consist of stars from the same progenitor of the H99 group, but stripped at an earlier time, before dynamical friction dragged in the satellite more closely. In total, based on their $(L_z,L_\perp)$-distributions, \citeauthor{refi05} counted 16 stars out of 410 (3.9\%) as putative members of the H99 streams, consistent with \citeauthor{gou03}'s \citeyear{gou03} estimate that no single stream could contain more than 5\% of the local halo stars\footnote{Even more so, since it could be expected that the 16 stars from \citeauthor{refi05} are distributed between three H99 streams, i.e., the kinematic group consisting of 3 stars plus the original two streams with positive and negative $W$ velocities, respectively.}. In this respect, it seems that the H99 progenitor is one of the major contributors to the local halo -- a result that has further been supported when \citet{kep07} analysed a sample of 231 nearby ($d\leq2.5$ kpc) metal-poor red giants, RR Lyrae and RHB stars, including again a subset of the \citet{bee00} catalog. These authors identified 11 or possibly 12 stars connected to the H99 group ($\sim5\%$) and showed that their progenitor was likely accreted between 6 and 9 Gyr ago by comparing the observed asymmetric $W$-distribution of the stars (8/11 with negative $W$) to simulations. In addition, [$\alpha$/Fe] ratios for three of the H99 stars revealed that they are very similar to other solar neighborhood halo stars, implying formation in a massive progenitor where enrichment occured mainly through SNe~II. 

The chemical similarity of the stream stars among each other, but also with normal inner halo stars, was recently shown by \citet{roe10} through determinations of abundances or upper limits, respectively, for 46 elements in 12 kinematically selected member stars from the lists of \citet{chi00}, \citet{refi05} and \citet{kep07}. \citeauthor{roe10} found that the stream stars span a broad range in [Fe/H] from -3.4 to -1.5, but compared to typical present-day dSph and ultra-faint dwarf galaxies, only small star-to-star dispersions of [X/Fe] at a given metallicity have been found. The high $\alpha$ abundances (e.g. [Ca/Fe]=+0.4) together with the trend of the $n$-capture abundances (e.g. [Ba/Fe] and [Eu/Fe]) to increase with metallicity point towards a scenario in which star formation in the H99 progenitor was truncated before SNeIa or AGB stars enriched the ISM.

Four members of the H99 stream with negative $W$ velocities also showed up when \citet{det07} re-analyzed the \citet{bee00} data using \citeauthor{dek76}'s \citeyear{dek76} theory of Galactic orbits and searching for streams in the space spanned by $(\nu,V_\text{az},V_{\Delta\text{E}})$ (see \S~\ref{sec:s31}). The same search strategy was used by \citet{kle09} on a sample of 22,321 nearby ($d\leq2$ kpc) subdwarfs with [Fe/H]<-0.5 from the seventh SDSS data release with astrophysical parameters from SEGUE \citep{yan09}. This sample, being fainter than the \citeauthor{bee00} compilation, added another 21 stars to the H99 stream. Referred to the total number of likely halo stars in the SDSS sample ($\sim4400$), this only accounts for 0.5\% of the local halo. I should note that the search strategy only picked out these 21 stars from the \textit{single} stream with negative $W$-velocities, similar to \citeauthor{det07}'s work from \citeyear{det07}. If one assumes the asymmetry in the $W$-distribution in the sample of \citet{kep07} as representative, another $3/11\cdot21\approx6$ H99 stars with positive $W$ velocities could be expected, presumably too few to appear in the significance maps of \citet{kle09}. \citet{kep07} also found that 67\% of their H99 member stars lay within $d=1$ kpc, a range which is incompletely sampled in \citet[][Fig.~5c]{kle09}. On the other hand, \citet{smi09} predict up to $\sim7\%$ of all halo stars within 2.5 kpc to belong to the H99 streams. Their result is based on the fraction of 12 putative H99 members out of 645 SDSS subdwarfs selected from a reduced proper motion diagram, again using the $(L_z,L_\perp)$ criterion and correcting for any selection effects due to the reduced proper motion bias. An explanation for these different results could be that the method of \citet{det07} and \citet{kle09} identifies single streams from the same progenitor in different regions of $(\nu,V_\text{az},V_{\Delta\text{E}})$-space, requiring a large number of stars in any single stream for a significant detection; in contrast, these single streams all enhance the signal in the same region of the $(L_z,L_\perp)$-plane.  

\subsubsection{$\omega$\,Centauri, the Kapteyn and Arcturus streams\label{sec:s413}}
\begin{figure}[ht]\sidecaption
	\centering
		\includegraphics[width=0.45\textwidth]{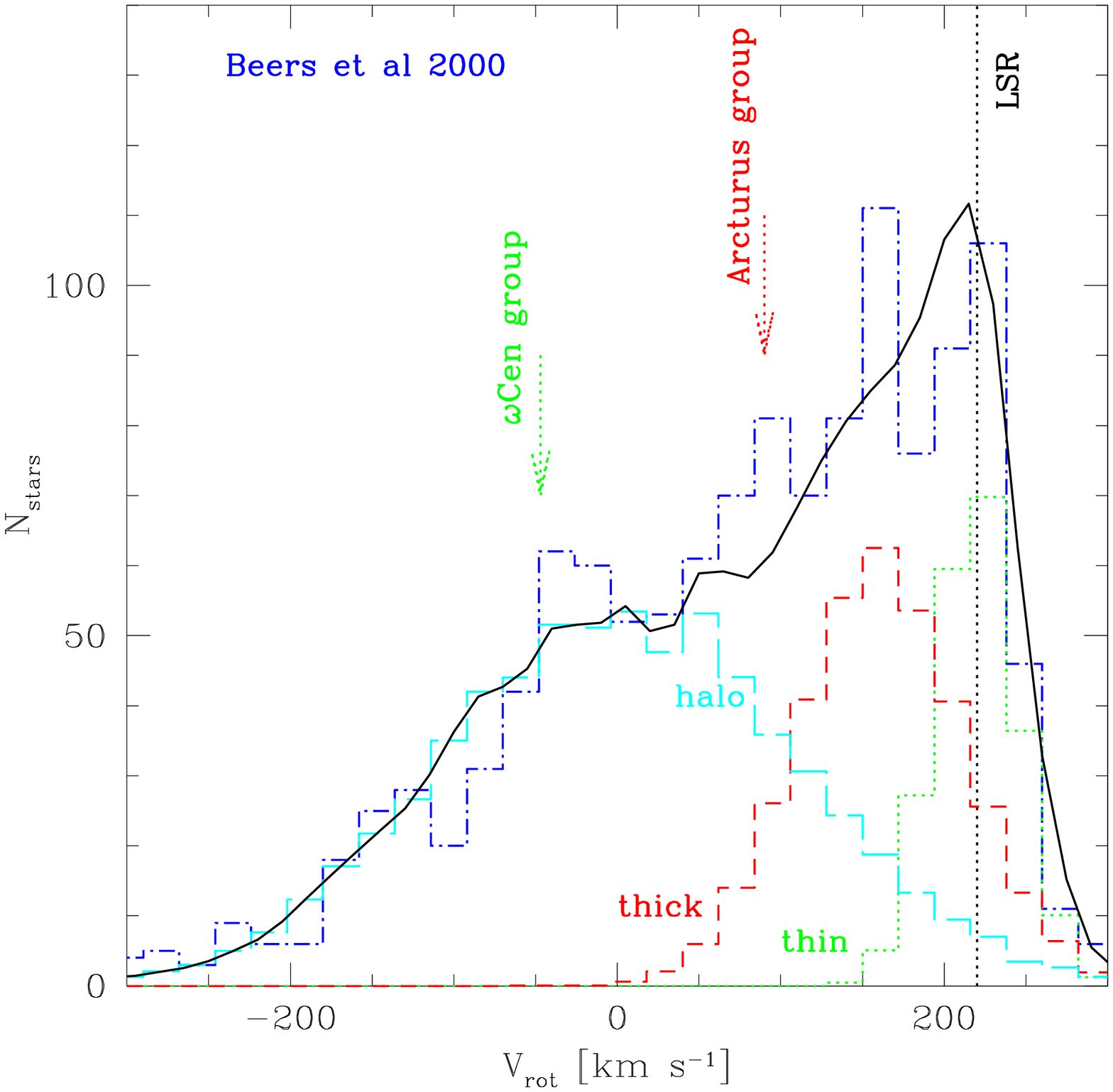}
	\caption{Histogram of rotational velocities $V_\text{rot}$ for all stars from the \citet{bee00} catalog of metal-poor stars (dashed histogram). The distribution is decomposed into three Gaussian components representing the thin disk, thick disk and halo as labeled. Note the excess of stars lagging the LSR with $\sim120$ km s$^{-1}$ (Arcturus stream) and $\sim270$ km s$^{-1}$ (\ocen stream).}
	\label{fig:f6}
\end{figure}

The \citet{bee00} catalog, besides containing the H99 stream as the most significant feature, reveals further substructure when plotting a histogram of the rotational velocities of the stars like in Figure~\ref{fig:f6}. The peak of stars with rotational velocities of $V_\text{rot}\sim100$ km s$^{-1}$ was investigated in detail by \citet{nav04} and attributed to the Arcturus stream, first described in 1971 by Eggen \citep{egg71} as a moving group of stars with similar kinematics as the star Arcturus. Originally, \citet{nav04} suggested a tidal origin for the Arcturus moving group, supported by the well-defined sequence of abundance ratios of member stars taken from the then available \citet{gra03} catalog. \citet{ari06}, who later detected the Arcturus group in a compilation of various catalogs using their $(V_\text{az},V_{\Delta\text{E}})$ method, suggested an external origin, too, based on the goodness of theoretical 12 Gyr isochrone fits to the putative member stars. The same suggestion was made by \citet{helm06} for a group of stars (their Group 2) detected in the APL-space distribution of Geneva-Copenhagen survey stars \citep{nord04} and possibly related to the Arcturus stream (according to its $V$- and [Fe/H]-distribution).

However, recent detailed abundance studies including various $\alpha$ and other elements throw doubt on the hypothesis that the Arcturus stream is composed of a homogeneous stellar population \citep{wil09}. Instead, the putative members\footnote{The \citet{wil09} selection of Arcturus group members by $V=-100\pm10$ km s$^{-1}$ differs slightly from the definition of the stream in other works: $\langle V\rangle=-110$ km s$^{-1}$ \citep{egg96}; $\langle V\rangle=-120$ km s$^{-1}$ \citep{nav04}; $\langle V\rangle=-125$ km s$^{-1}$ \citep{ari06}; $\langle V\rangle=-105$ km s$^{-1}$ \citep{bovy09a}. Note that such offsets could arise through different values for the LSR.} do not differ from the surrounding disk stars. This either indicates a progenitor system that had to be very massive to self-enrich to [Fe/H]=-0.6 or, more likely, a dynamical origin of this group. Similar to the Hercules stream, this could be either caused by the 6:1 outer Lindblad resonance with the bar at the solar position (B.~Fuchs, unpublished) or, as suggested recently by \citet{min09}, the relaxation of disk stars after the perturbation through a massive merger $\sim1.9$ Gyr ago. \citet{bovy09a} also suggest a dynamical origin due the very small $\sigma_V$ velocity dispersion they find for the Arcturus stream.

The overdensity of stars on slightly retrograde orbits in Figure~\ref{fig:f6} ($V_\text{rot}\sim-50$ km s$^{-1}$ or, equivalently, $V\sim-270$ km s$^{-1}$) was first noticed by \citet{din02}, most prominently when the sample was restricted to the metallicity range $-2.0\leq\text{[Fe/H]}\leq-1.5$. As these values are consistent with those of the peculiar globular cluster \ocen, which is confined to the disk ($z_\text{max}\sim1$ kpc) and has therefore long been suspected to be the surviving nucleus of a tidally disrupted dwarf galaxy \citep{free93}, \citeauthor{din02} hypothesized that the peak at $V_\text{rot}\sim-50$ km s$^{-1}$ was caused by tidal debris from that same galaxy \citep[see also][]{brook04a}. Further support for this theory came from \citet{meza05}, who picked out 13 likely \ocen group members by their $L_z$ and $\vert W\vert$ values from the compilation of $\sim150$ nearby metal-poor stars from \citet{gra03}. The broad ($-300$ km s$^{-1}\lesssim U\lesssim300$ km s$^{-1}$), symmetric and centrally under-represented $U$-distribution of these stars indicates that most of them currently move away and towards their apocenter, with only two stars having $\vert U\vert<50$ km s$^{-1}$. This agrees well with the apocenter of \ocen currently lying inside the solar circle at $\sim6$ kpc. Apart from the two kinematic outliers, all stars additionally define a narrow track in the ([$\alpha$/Fe],[Fe/H])-plane,consistent with the chemical composition of \ocen itself and indicative of a protracted phase of star formation in a self-enriching progenitor system. With $\sim7\%$, the \ocen group contributes to the local halo comparable to the H99 stream \citep{meza05}.

\ocen's kinematics also put it in the retrograde region occupied by the Kapteyn moving group, named so by O.~J. Eggen after Kaptey's star which lags the LSR with $\approx290$ km s$^{-1}$ \citep[Table~11 in][and references therein]{egg96}. Recently, 16 members of this stream listed by \citeauthor{egg96} have been observed spectroscopically to test the hypothesis that Kapteyn's group in fact originates from the same progenitor as \ocen \citep{wyl09}. At least 14 stars have been shown to be kinematically and chemically ([$\alpha$/Fe] plus [Cu/Fe] for 5 stars) consistent with \ocen's stellar populations, but due to their higher energies unlikely to stem from the globular cluster itself. Instead, the possible division of the stars into three $L_z$ bands made \citeauthor{wyl09} hypothesize that these stars were stripped from \ocen's parent galaxy at different orbital passages. The progenitor itself would have had a luminosity of $M_V\sim-11$ or $2\times10^6 L_\odot$ and a total mass of a few $10^8 M_\odot$. Interestingly, \citeauthor{wyl09} estimated that such a satellite galaxy is expected to shed several hundred more stars into the solar neighborhood that have yet to be detected. In this regard the retrograde streams found recently by various authors \citep[][see also Table~\ref{tab:t1}]{det07,kle09} could add further evidence for tidal debris from \ocen's progenitor system.  

\subsubsection{Further halo substructure in the solar neighborhood\label{sec:s414}}
Besides the H99 and \ocen streams, which seem to dominate the fraction of nearby halo stars attributed to tidal debris, the recent years have added several candidates for further halo streams thanks to the increasing number of large samples like SDSS/SEGUE. \citet{ari06} noted an overdensity of stars with thick disk kinematics rotating approximately 20 km s$^{-1}$ faster than the Arcturus group, but with very similar metallicity and age (inferred from theoretical isochrones). They confirmed this detection in the Geneva-Copenhagen catalog of solar neighborhood stars. Interestingly, this catalog was also analyzed by \citet{helm06}, who noted an overdensity of stars in the APL space in an eccentricity range of $0.3\leq e<0.5$, with overall kinematics overlapping with the detections of Arcturus and the new stream found by \citet{ari06}. \citeauthor{helm06} further decomposed the stars in this eccentricity range into three groups, based on the trend of their metallicity to decrease with increasing eccentricity in a discontinous fashion. Based on the metallicity distributions, they concluded that the three groups may well be tidal debris from three different progenitors with initial masses of $\sim4\times10^8$ M$_\odot$. It should be noted that the $(U,V)$-distributions of these stars indeed show the opposite trend of what would be expected if they had a dynamical (resonant) origin.

The model of \citet{min09}, where the streams found by \citeauthor{ari06} result from the relaxation of a perturbed disk following a major merger, is also able to explain the location of the moving group detected by \citet{kle08} in the first RAVE public data release \citep{stei06}, which is at $V\approx-160$ km s$^{-1}$. The original detection of the stream in RAVE data was based on the assumption that the vast majority of RAVE stars are dwarfs/~subdwarfs in order to estimate distances through the photometric parallax method. This assumption was shown to be doubtful; in fact, \citet{sea08} derived from reduced proper motions that $\sim44\%$ of RAVE stars are likely K-M giants. The now available second data release \citep{zwi08}, which contains spectroscopically derived $\log g$ estimates for $\sim21,000$ stars, in principal allows the classification of dwarfs and (sub-)giants and a critical re-analysis of a pure dwarf sample (Klement et al. 2010, in preperation).  
However, evidence for the existence of the KFR08 stream was found by \citet{kle09} in their sample of SEGUE subdwarfs at a confidence level of $\sim99.7\%$. Furthermore, \citet{bob10} identified 19 candidates of the KFR08 stream among a sample of F and G stars for which accurate distance estimates ($\sigma_\pi/\pi<0.15$) were available from the updated \textit{Hipparcos} catalog \citep{van07}; they derived a stream age of 13 Gyr from stellar isochrones and a mean metalliciy of $\langle\text{[Fe/H]}\rangle=-0.7$ with standard deviation 0.3 dex from Str{\"o}mgren $uvby\beta$ photometry. The chemical and chronological homogeinity, as well as the banana-shaped $(U,V)$-distribution of the stars \citep[Fig.~8 in][]{bob10}, favor a tidal rather than a dynamic origin of the KFR08 stream.

Other moving group candidates without known progenitor systems have been identified by \citet{det07} and \citet{kle09} in the space of effective integrals of motion $(\nu,V_\text{az},V_{\Delta\text{E}})$ (\S~\ref{sec:s31}). They are listed in Table~1 in the next section. One of the features detected by \citeauthor[][their `S$_3$' overdensity,]{det07} was confirmed by \citeauthor{kle09} and affiliated to a much larger group of stars that also show a spatial coherence with similarity to an overdensity detected by \citet{jur08} at $(R,z)\approx(9.5,0.8)$ kpc. The stream members move on slightly retrograde orbits, are currently near their apocenters and have a [Fe/H]-distribution characterized through $\langle\text{Fe/H]}\rangle\approx-1.6$, $\sigma_\text{[Fe/H]}=0.4$. Although this points towards a possible relation with \ocen, the high $W$ velocity components and spatial concentration of the stars argue against this hypotheses. A detailed chemical study would be helpful in solving this issue.

\subsection{A census on the currently known halo streams} 
In Table~\ref{tab:t1}, I summarize all currently identified halo streams within 2 kpc from the sun, ordered by the $v_\phi$ velocity components or, equivalently, their $L_z$ values. $v_\phi$ is the rotational velocity in a cylindrical coordinate system and for nearby stars can be approximated by $v_\phi\approx V+V_\text{LSR}+V_\odot$. I kept the names originally given to the individual streams by the authors cited at the end of each row. I have omitted to list Arcturus and the AF06 stream \citep{ari06}, because their kinematics and chemistry more closely resemble dynamical thick disk streams \citep{wil09}.  

From Table~\ref{tab:t1} it is obvious that some of the individual groups might be related in the sense that they might stem from the same progenitor. For example, many of the retrograde streams in the region $-100 \text{km s}^{-1}\lesssim v_\phi\lesssim0$ overlap in their rotational velocity and [Fe/H] abundances. As discussed in \S~\ref{sec:s413}, this region is thought to be populated by multiple streams from \ocen's progenitor, and it seems very likely that many of these features are indeed related. In particular, \citet{kle09} already showed that the streams `C2', `C3' and `S$_3$' are not only kinematically and chemically connected, but also spatially coherent, similar to a small overdensity at $(R,z)\approx(9.5,0.8)$ kpc described by \citet{jur08}. Other streams with a common origin might be the RHLS and H99 groups \citep{refi05}, R2 and KFR08 \citep{kle09} and the S$_3$ streams from \citet{det07} and \citet{kle09}, respectively. 

Furthermore, relations between halo streams and certain globular clusters have been suggested in several instances, based on their common position in the IoM-space. In particular, relations have been suggested between SKOa and the clusters NGC 5466, NGC 6934, NGC 7089/\,M2 and NGC 6205/\,M13 \citep{smi09} or between the H99 stream and M15 \citep{roe10}. Also, as already discussed in \S~\ref{sec:s413}, Kapteyn's stream has been linked to \ocen, and probably the globular clusters NGC 362 and NGC 6779 are associated, too \citep{din02}.

To draw further conclusions about the origins of, and relations between, the several streams one would need (i) more precise 3D velocity estimates \citep[$\lesssim$5 km s$^{-1}$][]{helm99b}, which will ultimately become available for thousands of halo stars from the anticipated \textit{Gaia} mission and (ii) detailed abundance analyses of the individual streams like the ones undertaken by \citet{wyl09} and \citet{roe10}. 

\section{Conclusions}
In this paper, my goal was to summarize the current state of research on stellar halo streams in the solar neighborhood. I gave an overview on the processes that play a role in the disruption of Milky Way satellites and the subsequent formation of stellar streams that eventually pass the vicinity of the sun. Disentangling and correlating the individual streams that contribute to the local halo thus gives valuable insights into the history of the Milky Way in the context of hierarchical galaxy formation. Various search strategies in complementary data sets should be used for this task in order to increase the confidence levels on the existence of individual streams. Further, spectroscopic studies of the various streams in the style of \citet{wil09}, \citet{wyl09} or \citet{roe10} are necessary to infer clues about their origin (dynamical vs. accreted), properties of any progenitor system and eventual relationships between individual streams. To fully resolve the hundreds of expected individual halo streams, 3D velocities out to at least 2 kpc have to be measured with accuracies better than 5 km s$^{-1}$. In addition, estimating their time since accretion requires precise measurements of parallaxes good to 1--2\% \citep{gom10b}. Ultimately, \textit{Gaia} will deliver these informations, but -- as this review hopefully demonstrated -- in the meantime, many discoveries and conclusions are already possible with currently available data and facilities.

\clearpage
\begin{deluxetable}{c|c|c|c|c|c|c}
\tabletypesize{\scriptsize}
\tablewidth{0pt}
\tablecaption{Current census of solar neighborhood halo streams\label{tab:t1}} 	
			\tablehead{\colhead{Stream}  & \colhead{$\langle v_\phi\rangle$} & \colhead{$\sigma_{v_\phi}$} & \colhead{$\langle\text{[Fe/H]}\rangle$} & \colhead{$\sigma_\text{[Fe/H]}$} & \colhead{$N$} & \colhead{References} \\
\colhead{}  & \colhead{(km s$^{-1}$)} & \colhead{(km s$^{-1}$)} & \colhead{} & \colhead{} & \colhead{} & \colhead{} }
\startdata
			C2  &   -75  &   24  &   -1.6  & 0.4  &  53  &  \citet{kle09}\\
			S$_3$  &  -48 \tablenotemark{a}  &  14   &  -1.6  &  0.4  &  33  &  \citet{kle09}\\
			Kapteyn  &  -46 \tablenotemark{b}  &  63  &  -1.5  & 0.3  &  14  &  \citet{wyl09}\\
			\ocen & -30  & \nodata & \nodata & \nodata & 56  &  \citet{din02}\\
			C3  &  -24  &  14  & -1.7  &  0.4 &  44  &  \citet{kle09}\\
			C1   &  -16    &  9   &  -1.5  &  0.2  &  32 & \citet{kle09}\\
			S$_3$  &  -9 \tablenotemark{a} & 10 & -1.9  & 0.5 & 10  & \citet{det07}\\
			S$_2$  &  -7  & 4 & -1.9  & 0.1 &  4 & \citet{det07}\\
			S$_1$  &  21  & 2 & -1.7  & 0.5 &   4 & \citet{det07}\\
			SKOa  &  43  &  25  &  -2.0  & 0.2  &  6  &  \citet{smi09}\\
			R2 &  59      &  15  &  -1.4  &  0.3  &  19 & \citet{kle09}\\
			KFR08 &  69  &  11  &  -0.7  &  0.3  &  19  & \citet{bob10}\\ 
			Groombridge 1830 & 71 \tablenotemark{c} & 18 & \nodata & \nodata &  5  &  \citet{egg59}\\
			RHLS  &  99  & 25  &  -2.0  &  0.2  &  3 & \citet{refi05}\\
			H99  &  140     &  33  &  -1.8  &  0.4  &  33 & \citet{kep07,kle09}\\
			C4  &  173  &  9   &  -2.3   & 0.3  &  20  &  \citet{kle09}\\     
	\enddata
 \tablecomments{$N$ is the number of total stream
members given in the references, $\langle v_\phi\rangle$ and $\langle\text{[Fe/H]}\rangle$ are the mean rotational velocity and metallicity of the stream, and
$\sigma_{v_\phi}$, $\sigma_\text{[Fe/H]}$ the corresponding standard deviations.}
\tablenotetext{a}{From the discrepancy between the $v_\phi$ velocities of the S$_3$ stream from \citet{det07} and \citet{kle09} it follows that both authors might have described two different, but maybe related streams.}   
\tablenotetext{b}{The $V$ values adopted by \citet{wyl09} are based on modern parallaxes, proper motions and radial velocities and differ from the more tightly confined $V$ values originally given in \citet{egg96}.}
\tablenotetext{c}{\citet{egg59} used $V_\odot=17$km s$^{-1}$, while I applied 5.2 km s$^{-1}$. Note that this group has not yet been confirmed by later studies.}
\end{deluxetable}

 \clearpage
\section*{Acknowledgements}
I would like to thank the following people: Burkhard Fuchs, Chao Liu and Amina Helmi for useful comments on an early draft of this paper; Christian Dettbarn for kindly providing me the parameters of the individual stars from his paper; Amina Helmi for allowing me to use the data of her simulations in Figures~\ref{fig:f2} and \ref{fig:f4}; Facundo A. Gomez and Andres Meza for providing me with high-resolution figures from their papers. I acknowledge financial support from \textit{Deutsches Zentrum f{\"u}r Luft- und Raumfahrt}.

\clearpage
\end{document}